\documentstyle[fullpage,graphics]{article}

\newcommand{\vecA}{{\bf A}}
\newcommand{\vecr}{{\bf r}}
\newcommand{\vecp}{{\bf p}}
\newcommand{\vecv}{{\bf v}}
\newcommand{\vecs}{{\bf s}}
\newcommand{\vecB}{{\bf B}}
\newcommand{\vecR}{{\bf R}}
\newcommand{\vecf}{{\bf f}}
\newcommand{\vecx}{{\bf x}}
\newcommand{\vecrdot}{{\dot{\vecr}}}
\newcommand{\vecnabla}{{\bf \nabla}}
\newcommand{\vecj}{{\bf j}}
\newcommand{\vecPi}{{\bf \Pi}}

\newcommand{\rhat}{{\hat{\vecr}}}
\newcommand{\phat}{{\hat{\vecp}}}
\newcommand{\Ahat}{{\hat{\vecA}}}
\newcommand{\Pihat}{{\hat{\vecPi}}}
\newcommand{\vhat}{{\hat{\vecv}}}
\newcommand{\What}{{\hat{W}}}
\newcommand{\Uhat}{{\hat{U}}}
\newcommand{\phihat}{{\hat{\phi}}}
\newcommand{\vnl}{{V_{nl}}}
\newcommand{\vnlop}{{\hat{V}_{nl}}}
\newcommand{\hop}{{\hat{H}}}
\newcommand{\Kloc}{{K_{loc}}}
\newcommand{\Knl}{{K_{nl}}}
\newcommand{\calH}{{\mathcal{H}}}
\newcommand{\calW}{{\mathcal{W}}}
\newcommand{\calN}{{\mathcal{N}}}

\title{Coupling of Nonlocal Potentials to Electromagnetic Fields}
\author{Sohrab Ismail-Beigi, Eric K. Chang, and Steven G. Louie\\
Department of Physics, University of California,
Berkeley, CA  94720, and\\
Materials Sciences Division, Lawrence Berkeley National Laboratory,
Berkeley, CA 94720}

\date{\today}

\begin{document}

\maketitle

\begin{abstract}
Nonlocal Hamiltonians are used widely in first-principles quantum
calculations; the nonlocality stems from eliminating undesired degrees
of freedom, e.g. core electrons.  To date, attempts to couple nonlocal
systems to external electromagnetic (EM) fields have been heuristic or
limited to weak or long wavelength fields.  Using Feynman path
integrals, we derive an exact, closed-form coupling of arbitrary EM
fields to nonlocal systems.  Our results justify and clarify the
couplings used to date and are essential for systematic computation of
linear and especially nonlinear response.
\end{abstract}

\vspace{0.5cm}

Theoretical study of the electronic response of quantum systems to
electromagnetic (EM) probes is central to understanding experimental
results and what they imply about the structure of matter.  It is now
possible to predict the response, from first principles, with
sufficient accuracy to compare quantitatively to experiment
\cite{book}.  Nonlocal pseudopotentials are a key ingredient of many
{\em ab initio} electronic structure calculations as they eliminate
the largely inert and thus physically unimportant core electrons.  The
price paid is that the valence electrons feel a nonlocal atomic
potential.

Hence, it is vital to know how electrons moving in nonlocal potentials
couple to EM fields.  Early work \cite{LuttingerKohn} found the
coupling within the effective-mass framework.  To date, refs.
\cite{Kolorev,Starace,almost} represent the only first principles
attempts to find the coupling, but with limited success.  In
ref. \cite{Kolorev}, the coupling terms involve an integral of the
field over an unspecified path.  Path ambiguity, discussed below, is a
central problem that leads to non unique couplings and incorrect
quantitative results.  Ref. \cite{Starace} stops at the formal level
due to operator ordering ambiguities, and an explicit, practical form
of the coupling is given only for the long wavelength (i.e. dipole)
approximation.  This treatment excludes the description of AC magnetic
fields or, more generally, spatially varying EM fields.  Operator
ordering problems also force the results of ref. \cite{almost} to be
limited to linear (i.e. weak) coupling.

Other attempts are also limited to linear coupling
\cite{dielecs,chimag} where physical intuition leads to replacing the
momentum by the velocity operator. When seeking nonlinear response,
the coupling is either considered only for longitudinal fields
\cite{LevineAllan,TDLDAnonlin} or only within the long wavelength
(dipole) approximation \cite{Girlanda}.

Our work presents a rigorous derivation of an explicit, closed-form
expression for the coupling of nonlocal systems to {\em arbitrary} EM
fields.  The value of this result is several fold: e.g., having the
correct coupling is essential for accurate and systematic {\em ab
initio} calculation of linear and nonlinear response.  Aside from
first principles methods, many semi-empirical or phenomenological
models use nonlocal Hamiltonians with tight-binding like hopping
terms.  There have been only heuristic attempts to couple such systems
to EM fields \cite{TB}.  Our results justify the assumed forms of
previous couplings.

Although {\em ab initio} calculations involve interacting electrons,
nonlocal potentials are {\em one-body} operators.  For clarity, we
derive our results for a one-electron system.  The results generalize
directly to the interacting case.

A main principle of electrodynamics is gauge-invariance.  A particle
in a local potential $U(\vecr)$ is governed by the Hamiltonian $
\hop_0(\rhat,\phat) = \phat^2/2m + \Uhat(\rhat) $.  The standard
prescription for the gauge-invariant coupling uses the minimal
substitution $\phat\rightarrow \Pihat\equiv \phat - \Ahat(\rhat,t)$
($\vecA$ is the vector potential). The coupled Hamiltonian is
$\hop_\vecA(\rhat,\phat) = \Pihat^2/2m + \Uhat(\rhat) +
q\phihat(\rhat,t)$ ($\phi$ is the electrostatic potential).  We use
units where $q/c=\hbar=1$ ($q$ is the particle's charge and $c$ the
speed of light).

The problem of interest, however, has a nonlocal potential $\vnl$ with
Hamiltonian
\begin{eqnarray}
\hop_0 & = & \phat^2/2m + \Uhat(\rhat) + \vnlop\,,
\label{eq:H0start}
\end{eqnarray}
where $\vnl(\vecr,\vecr')\equiv \langle\vecr|\vnlop|\vecr'\rangle\neq
0$ for $\vecr\neq\vecr'$.  Gauge-invariance demands that $\vnl$ depend
on $\vecA$.  A major obstacle to finding $\hop_\vecA$ is that
$\vnl(\vecr,\vecr')$ is not specified in terms of the canonical
operators $(\rhat,\phat)$.  Below, we first rewrite $\vnl$ in terms of
$(\rhat,\phat)$, and then perform minimal substitution in an
unambiguous manner to find $\hop_\vecA$.

Starting with $\vnlop$, we perform a Weyl transformation to obtain an
equivalent operator $\What$ of $(\rhat,\phat)$ \cite{Weyl}:
\begin{equation}
\What(\rhat,\phat) = \int \! d\vecs \, e^{-i\phat\cdot \vecs/2} \,
\vnl\left(\rhat+{\vecs\over 2},\rhat-{\vecs \over 2}\right) \, e^{-i\phat\cdot
\vecs/2}\,.
\label{eq:weyloptrans}
\end{equation}
It is straightforward to show that $\langle
\vecr|\What(\rhat,\phat)|\vecr'\rangle = \langle
\vecr|\vnlop|\vecr'\rangle$.  Thus $\vnlop$ and $\What$ are the {\em
same} operator. We now have written the Hamiltonian without fields as
\begin{equation}
\hop_0(\rhat,\phat) = \phat^2/2m + \Uhat(\rhat) + \What(\rhat,\phat)\,.
\label{eq:H0}
\end{equation}
For the equivalent classical system, $\calW(\vecr,\vecp)$ is a
function of the numbers $\vecr$ and $\vecp$ given by
\begin{equation}
\calW(\vecr,\vecp) = \int d\vecs \, \vnl(\vecr+\vecs/2,\vecr-\vecs/2) \,
e^{-i\vecp\cdot\vecs} \,.
\label{eq:Wdef}
\end{equation}

At this stage, we would like to employ the minimal substitution
$\phat\rightarrow\Pihat$ in Eq.~(\ref{eq:H0}) to find the coupled
Hamiltonian $\hop_\vecA$.  Unfortunately, the components of $\Pihat$
do not commute so that initially equivalent orderings of operators in
$\What$ lead to physically different couplings: e.g., $\phat_x\phat_y
= \phat_y\phat_x$ but $\Pihat_x\Pihat_y \neq \Pihat_y\Pihat_x$ since
$[\Pihat_x,\Pihat_y]=i\vecB_z$.  Continuing along this line requires
applying ordering postulates to the operators in $\What$ {\em before}
minimal substitution to guarantee unambiguous final results.  Only for
long wavelengths (i.e. spatially uniform $\vecA$) is the substitution
unambiguous, and the results of \cite{Starace,Girlanda} are limited to
this case.  Similarly, this same ordering issue forced the results in
\cite{almost} be limited to first order in $\vecA$.

An equivalent statement of the problem is that different evaluations
of $e^{-i\phat\cdot\vecs/2}$ in Eq.~(\ref{eq:weyloptrans}) correspond
to translating by $\vecs/2$ along different paths.  This path
ambiguity is present in \cite{Kolorev} where line integrals of $\vecA$
are used.  When $\vecA$ is constant, all paths give identical results.
But once $\vecA$ has spatial variation, different paths contribute
different phases leading to non unique differing couplings.

To find the solution, we first consider the classical system governed
by $\calW(\vecr,\vecp)$.  Minimal substitution is well defined
classically since $\vecp$ and $\vecr$ are commuting numbers.  The
classical coupled Hamiltonian is
\begin{eqnarray}
\calH_\vecA(\vecr,\vecp) = \vecPi^2/2m + U(\vecr) + \calW(\vecr,\vecPi) +
\phi(\vecr,t)\,.
\label{eq:Hclassical}
\end{eqnarray}
Next, the Feynman path-integral formulation provides the {\em quantum}
evolution of a system by summing over its {\em classical} dynamics.
The propagation amplitude $K$ between the two space-time points
$(\vecr',0)$ and $(\vecr,t)$ in the operator formulation is given by
\cite{NegeleOrland}
\begin{eqnarray}
K(\vecr,t;\vecr',0) & \equiv & \langle \vecr|T\exp\left(-i\int_{0}^{t}
d\tau\, \hop_\vecA(\tau)\right)|\vecr'\rangle\nonumber\\ & = &
\delta(\vecr-\vecr') - it\,\langle \vecr|\hop_\vecA(t)|\vecr'\rangle +
O(t^2)\,.
\label{eq:Kdef}
\end{eqnarray}
Now $K$ also equals the result of the Feynman formulation where we sum
over all classical trajectories in phase space $(\vecr,\vecp)$
weighted by each path's action $S$ \cite{Zagoskin}:
\begin{eqnarray}
K(\vecr,t;\vecr',0) & = & \int' D[\vecr] \int D[\vecp]
\,e^{iS}\label{eq:pathint}\,, \\  S & = & \int_0^t
d\tau\,\left[ \vecp(\tau)\cdot
\vecrdot(\tau)-\calH_\vecA(\vecr(\tau),\vecp(\tau))\right]\,.
\label{eq:action}
\end{eqnarray}
The prime on the $\vecr$ path integral means that we only sum over
paths where $\vecr(0)=\vecr'$ and $\vecr(t)=\vecr$.

We seek the quantum Hamiltonian $\hop_\vecA(t)$ which is the $O(t)$
term in Eq.~(\ref{eq:Kdef}).  Hence we will expand the path integral
of Eq.~(\ref{eq:pathint}) to $O(t)$ and extract $\hop_\vecA(t)$.

We now use the fact that $\calW(\vecr,\vecp)$ is generally bounded,
i.e. $|\calW(\vecr,\vecp)|<C$ for some number $C$.  This is true for
atomic pseudopotentials $\vnl(\vecr,\vecr')$ which are smooth and have
compact support.  Thus when $t\rightarrow 0$, the contribution of the
$\calW$ term to $S$ in Eq.~(\ref{eq:action}) is $O(t)$.  Hence we need
expand only to linear order in $\calW$:
\begin{eqnarray}
K & = & \Kloc + \Knl\nonumber\,,\\ \Kloc & = & \int' D[\vecr] \int
D[\vecp]\, e^{iS_{loc}}\nonumber\,,\\ \Knl & = & -i \int' D[\vecr]
\int D[\vecp] \int d\tau\,
\calW(\vecr,\vecPi)\,e^{iS_{loc}}\,,\nonumber\\ S_{loc} & = & \int
d\tau'\, \left[\vecp\cdot\vecrdot - \vecPi^2/2m - U(\vecr) -
\phi(\vecr,\tau')\right]\,.
\end{eqnarray}
The local propagator $\Kloc$ describes the textbook situation of a
local potential and is given by
\begin{equation}
\Kloc = \delta(\vecr-\vecr') -it\,\langle
\vecr|\hop_{loc}|\vecr'\rangle + O(t^2)\,,
\label{eq:Kloc}
\end{equation}
where $\hop_{loc} = \Pihat^2/2m + \Uhat + q\phihat$.  Thus, we
concentrate on the nonlocal propagator $\Knl$.

The local potential $U+q\phi$ depends only on $\vecr(\tau)$ and $\tau$
and not on $\vecp$.  Hence as $t\rightarrow 0$, the contribution to
$S_{loc}$ from $U+q\phi$ is $O(t)$.  We ignore this contribution to
$\Knl$ since it leads to terms of $O(t^2)$. However, $p$ scales as
$\vecp\sim(\vecr-\vecr')/t=O(1/t)$ so expanding in $t$ will fail for
terms involving $\vecp$.  Thus we must perform the $\vecp$ integral
explicitly.  At fixed $\vecr$, we change variables from $\vecp$ to
$\vecPi$:
\begin{eqnarray}
\Knl & = & -i \int' D[\vecr] \, e^{i\int d\tau'
\vecA\cdot\vecrdot}\ \Xi\ ,\\ \Xi & \equiv & \int D[\vecPi] \int
d\tau\, \calW(\vecr(\tau),\vecPi(\tau))\, e^{i\int
d\tau'[\vecPi\cdot\vecrdot-\vecPi^2/2m]}\nonumber\\ & = & \int d\tau\,
W\left(\vecr(\tau),{-i\delta\over\delta\vecrdot(\tau)}\right) \int D[\vecPi]
e^{i\int d\tau'[\vecPi\cdot\vecrdot-\vecPi^2/2m]} \nonumber\\ & = & \calN \int
d\tau\, W\left(\vecr(\tau),{-i\delta\over\delta\vecrdot(\tau)}\right)
e^{im/2\int d\tau'\vecrdot^2}\,.
\end{eqnarray}
Above, we use a functional derivative to pull $\calW$ out of the
$\vecPi$ integral.  This allows us to perform the Gaussian $\vecPi$
integral, which yields a normalization constant $\calN$.

Using the definition of $\calW$ from Eq.~(\ref{eq:Wdef}), we have
\begin{eqnarray}
W\left(\vecr,{-i\delta\over\delta\vecrdot}\right) & = & \int
d\vecs\,\vnl\left(\vecr+{\vecs\over2},\vecr-{\vecs\over2}\right)
e^{-\vecs\cdot{\delta\over\delta\vecrdot}}\,.
\end{eqnarray}
The exponentiated functional derivative translates $\vecrdot$:
\begin{eqnarray}
\Knl & = & -i \calN \int \! d\tau \int \! d\vecs \int' \! D[\vecr]
\vnl\left(\vecr(\tau)+{\vecs\over2},\vecr(\tau)-{\vecs\over2}\right)
\times \nonumber\\ & & e^{i\int
d\tau'\left(m[\vecrdot-\vecs\delta(\tau'-\tau)]^2/2
+\vecA\cdot\vecrdot\right)}
\label{eq:kinkpath}
\end{eqnarray}
Note that changing the order of the integrals does not change the
final result.  We now change variables, at fixed $\vecs$ and $\tau$, from
$\vecr(\tau)$ to $\vecf(\tau)$ as follows:
\begin{equation}
\vecr(\tau') = \vecf(\tau') +
[\vecr'+\vecs\,\theta(\tau'-\tau)+(\vecr-\vecr'-\vecs)\tau'/t]\,.
\label{eq:xidef}
\end{equation}
where $d\theta(x)/dx=\delta(x)$.  This choice follows from the action
of $\vnl$ in Eq.~(\ref{eq:kinkpath}): $\vnl$ causes a transition from
$\vecr(\tau)-\vecs/2$ to $\vecr(\tau)+\vecs/2$ at time $\tau$.  We
parameterize the paths as fluctuations $\vecf$ about a straight-line
path with a discontinuity of size $\vecs$ at time $\tau$ (term in
brackets).  With this choice, $\vecf$ obeys the conditions
$\vecf(0)=\vecf(t)=0$.

Upon substituting $\vecf$, we have
\begin{eqnarray*}
\Knl & = & -i\calN \!\int \!d\tau\!\int \!d\vecs\!\int \!D[\vecf]
\vnl\left(\vecr(\tau)+{\vecs\over2},\vecr(\tau)-{\vecs\over2}\right)\times\\
& & e^{im/2(\vecr-\vecr'-\vecs)^2/t+
i\int d\tau'\left(m\dot\vecf^2/2+\vecA\cdot\vecrdot\right)}
\end{eqnarray*}
The conditions on $\vecf$ permit the Fourier expansion $\vecf(\tau') =
\sum_n \vecf_n \sin(\pi n \tau'/t)$.  The kinetic integral is
$\int_{0}^{t} d\tau'\dot\vecf^2 = \pi^2/2 \sum_n n^2\vecf_n^2/t$.  We
now note that for any analytic function $g(x)$, $\int dx\,
g(x)e^{ix^2/t} = \sqrt{i\pi t}[g(0)+O(t^2)]$, so that within an
integral we may set $e^{ix^2/t}=\sqrt{i\pi t}[\delta(x) + O(t^2)]$.
Since $\Knl$ is already $O(t)$ and we ignore terms of $O(t^2)$, we
replace $e^{ix^2/t}$ by $\sqrt{i\pi t}\delta(x)$ everywhere in the
above integral: this applies to the $\vecf_n^2/t$ as well as to the
$(\vecr-\vecr'-\vecs)^2/t$ terms.  Physically, fluctuations in $\vecf$
and $\vecs$ contribute a wildly oscillating large phase as
$t\rightarrow 0$ so only the stationary phase contribution at
$\vecf=0$ and $\vecs=\vecr-\vecr'$ remains when $t\rightarrow 0$.

Therefore, we set $\vecf=0$, $\vecs=\vecr-\vecr'$, absorb all
constants from the integrations into $\calN'$, and arrive at
\begin{eqnarray}
K_{nl} & = & -i\calN'\int \! d\tau\, e^{i\int d\tau'\,
\vecA\cdot\dot{\vecx}}\times\nonumber\\
& & \vnl\left(\vecx(\tau,\tau)+{\vecr-\vecr'\over2},
\vecx(\tau,\tau)-{\vecr-\vecr'\over2}\right)
\,,\label{eq:almost}  \\
\vecx(\tau,\tau')  & \equiv & 
\vecr(\tau')|_{\vecf=0,\vecs=\vecr-\vecr'} = \vecr' +
(\vecr-\vecr')\,\theta(\tau'-\tau)\nonumber\,.
\end{eqnarray}
Since we defined $d\theta(x)/dx \equiv \delta(x)$ and
$\delta(-x)=\delta(x)$, we have $\theta(0)=1/2$ so
$\vecx(\tau,\tau)=\vecr'+(\vecr-\vecr')/2$.

The final hurdle involves evaluating $\int d\tau' \vecA\cdot\dot\vecx$
in Eq.~(\ref{eq:almost}).  We replace the $\delta(x)$ appearing in
$\dot\vecx$ by any smooth, non-negative function $g(x)$ of compact
support about $x=0$ with width much smaller than $t$ and unit
integral.  We then replace $\theta(x)$ by $G(x)$ where
$dG(x)/dx=g(x)$, so $G$ increases monotonically from 0 to 1 in a width
much smaller than $t$.  The integral is then
\begin{eqnarray}
\int_0^t \!\! d\tau' \vecA(\vecx(\tau,\tau'),\tau')\cdot
\dot\vecx(\tau,\tau') & = & \int_{\vecr'}^\vecr \!\! \vecA(\vecx,t)\cdot
d\vecx + O(t^2)\nonumber\,.
\end{eqnarray}
We evaluate $\vecA$ at $t$ since any explicit time dependence of
$\vecA$ changes $\Knl$ by $O(t^2)$, which is ignored.  The integral is
then just along a {\em straight line} from $\vecr'$ to $\vecr$.

After replacing the $\tau$ integral by $t$ in Eq.~(\ref{eq:almost}),
our nonlocal propagator is $\Knl = -i t \, \calN' \,
\vnl(\vecr,\vecr') \exp(i\int_{\vecr'}^{\vecr}\vecA\cdot
d\vecx)+O(t^2)$. Recovery of the correct propagator for $\vecA=0$
requires $\calN'=1$.  Combining $\Knl$ with $K_{loc}$
(Eq.~(\ref{eq:Kloc})), we have as our central result the following
gauge-invariant quantum Hamiltonian
\begin{eqnarray}
\hop_\vecA & = &
(\phat-q\Ahat/c)^2/2m+\Uhat+q\phihat+\vnlop^\vecA
\,,\label{eq:mainresult1}\\ \langle \vecr|\vnlop^\vecA| \vecr'\rangle
& = & \vnl(\vecr,\vecr')\,e^{{iq\over\hbar
c}\int_{\vecr'}^{\vecr}\vecA(\vecx,t)\cdot d\vecx}\ \
\mbox{(straight-line)}.\label{eq:mainresult2}
\end{eqnarray}
In practice, we often use perturbation theory to compute response
functions: we set $\hop_A=\hop_0+\hop^{int}$ and work to a desired
order in the interaction $\hop^{int}$.  We expand the above result to
all orders in $\vecA$:
\begin{eqnarray}
\hop_\vecA = \hop_0 + q\phihat -{q(\phat\cdot\Ahat +
\Ahat\cdot\phat)\over 2mc} + {q^2\Ahat^2\over 2mc^2} +
\vnlop^{int}\,,\nonumber\\ \langle \vecr|\vnlop^{int}|\vecr'\rangle =
\vnl(\vecr,\vecr') \sum_{k=1}^\infty {1\over k!}\left({iq\over\hbar
c}\int_{\vecr'}^{\vecr} \vecA\cdot d\vecx\right)^k\label{eq:perturb}\,.
\end{eqnarray}

In relation to previous work, a coupling similar to
Eq.~(\ref{eq:mainresult2}) is given in Ref. \cite{Kolorev} but only
after the straight line path is assumed with no justification.
Refs. \cite{Starace} and \cite{Girlanda} both find a result equivalent
to the special case of Eq.~(\ref{eq:mainresult2}) in the limit of
spatially constant $\vecA$.  In Ref. \cite{almost}, the linear term
($k=1$) of Eq.~(\ref{eq:perturb}) is derived correctly, but attempts
at finding a general expression for arbitrary $\vecA$ are stymied by
operator ordering problems.

\begin{figure}
\resizebox{3.9in}{!}{\includegraphics{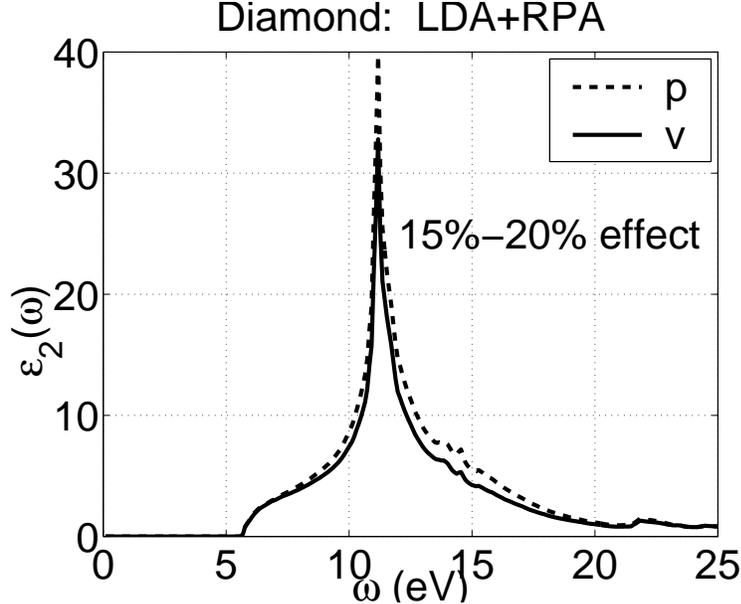}}
\caption{Absorption spectrum for diamond: $\varepsilon_2(\omega)$
vs. $\omega$ in eV.  The dashed curve uses the momentum operator, the
solid curve uses the velocity operator.}
\label{fig:velocity}
\end{figure}

We now apply our result to the case of linear optical response where
the wavelength of $\vecA$ is much greater than atomic dimensions and
we set $\vecA$ constant over the nonlocal range of $\vnl$.  For the
Coulomb gauge $\vecnabla\cdot\vecA=\phi=0$, the linear coupling is
($k=1$ term of Eq.~(\ref{eq:perturb}))
\begin{eqnarray}
\hop_1^{int} & = & -\Ahat(t)\cdot q\vhat/c\ \ \mbox{where} \ \
\vhat \equiv -i[\rhat,\hop_0]/\hbar\,.
\label{eq:hint1velop}
\end{eqnarray}
Here, $\vhat=d\rhat/dt$ is the velocity operator.  This result
justifies the intuition that $\vecA$ couples to the current
$\vecj=q\vecv$. For a local system $\vecv=\vecp/m$ as expected.  The
difference between velocity and momentum has been stressed when
computing optical response
\cite{dielecs,LevineAllan,Girlanda,nlandgauge}.  However, those works
assume the long wavelength limit or couple only to longitudinal
fields.  Our result is a direct demonstration that in the Coulomb
gauge, the coupling of electrons to long waves is via the velocity
operator.  To show the significance of nonlocality, we perform {\em ab
initio} pseudopotential density functional calculations \cite{RMP}.
Figure~\ref{fig:velocity} shows the RPA absorption spectrum,
$\varepsilon_2(\omega)$ with a 0.2 eV Lorentzian broadening, where
either the momentum or velocity operator is employed.  The effect of
nonlocality is 15\% to 20\%, non negligible when aiming for
quantitative comparison to experiment.

Next, we study electrons in a uniform magnetic field.  Magnetic fields
provide a novel application of our result since $\vecA$ must vary
spatially so that the usual long wavelength approximation is invalid.
We calculate the magnetic susceptibility $\chi=
-\partial^2E/\partial^2B$ of atomic carbon and neon.  We choose $\vecA
= -B(0,0,x+a)$: the gauge center is offset from the ionic nucleus to
simulate the realistic case when many atoms are present.  We choose
$a=4$ a.u. as a reasonable value.  We use $s$ and $p$
Kleinmann-Bylander nonlocal projectors \cite{RMP} ($p$ local) and
compute $\chi$ following ref. \cite{chimag}.  Calculating $\chi$
requires the linear and quadratic couplings, both of which have
nonlocal terms.  For linear coupling, using $\vhat$ instead of
$\phat/m$ has been advocated \cite{chimag}, whereas the quadratic
nonlocal term of Eq.~(\ref{eq:perturb}) is truly novel.  Table
\ref{table:chi} shows results for $\chi$ when using various possible
couplings: (1) the traditional local coupling $(\phat-\Ahat)^2/2m$;
(2) the linear coupling is from Eq.~(\ref{eq:perturb}) but the
quadratic coupling is the local $\Ahat^2/2m$; (3) both linear and
quadratic couplings are from Eq.~(\ref{eq:perturb}); (4) all-electron
results excluding the contribution of the core $1s$ states.  Method
(1) is qualitatively incorrect as it is not gauge-invariant.  Using
the correct linear coupling (method 2) greatly improves the quality of
the results.  However, the inclusion of our novel nonlocal quadratic
terms (method 3) yields $\chi$ in excellent quantitative agreement
with the desired all-electron results.

\begin{table}
\begin{center}
\begin{tabular}{c||c|c|c|c|c}
\ \ $-\chi$ \ \ & \ \ \ \ 1 \ \ \ \ & \ \ \ 2 \ \ \ & \ \ \ 3 \ \ \ & \ \ all-elec.\ \ & \ expt\cite{chimag}\  \\
\hline
\hline
C  & -40.1  & 16.9  & 12.80  & 12.85  & -- \\
\hline
Ne & -62.4  & 6.15   &  7.76  & 7.75   &  7.2
\end{tabular}
\end{center}
\caption{Negative of valence atomic magnetic suceptibility $-\chi$ in
cm$^3$/mole.  See text for details.}
\label{table:chi}
\end{table}

We emphasize that Eq.~(\ref{eq:mainresult2}) gives the coupling to any
order in $\vecA$.  This allows, for the first time, for direct and
{\em systematic} calculation of high-order nonlinear responses without
resort to longitudinal-only couplings \cite{LevineAllan} or to the
long wave limit for the transverse case
\cite{TDLDAnonlin,Girlanda,TDLDAreal}.

Finally, our result bears directly on systems with generic nonlocal
hopping terms, e.g. tight-binding systems, where electrons hop from a
localized state $\beta$ on site $\vecR'$ to a state $\alpha$ at
$\vecR$ with amplitude $t_{\alpha\beta}(\vecR,\vecR')$.  Our result
shows that, in the extreme tight-binding limit, the EM coupling
modifies the amplitude to become $t_{\alpha\beta}(\vecR,\vecR')
\exp\left({iq\over\hbar c} \int_{\vecR'}^{\vecR} \vecA\cdot
d\vecx\right)$.  This justifies the straight-line integral used in the
Peierls substitution and does not suffer from path ambiguity
\cite{TB}.

This work was supported by NSF grant \#DMR-0087088 and by the Office
of Energy Research, Office of Basic Energy Sciences, Materials Science
Division of the U.S. Department of Energy contract
\#DE-AC03-76SF00098.  We thank Kevin Mitchell and Andrew Charman for
illuminating the importance of Weyl transformations.  Young-Gui Yoon
provided helpful criticism of the ideas leading to this work.
Computer time was provided by the DOE at the Lawrence Berkeley
National Laboratory's NERSC center.

% now the references. delete or change fake bibitem. delete next three
%   lines and directly read in your .bbl file if you use bibtex.

\end{document}